\documentclass[aps,prl,twocolumn,amsmath,amssymb,superscriptaddress,showpacs]{revtex4}
\usepackage{graphicx}
\usepackage[latin1]{inputenc}
\usepackage{amsmath}

\begin{document}

\title{From electronic structure to catalytic activity: A single
  descriptor for adsorption and reactivity on transition-metal
  carbides}  

\author{A. Vojvodic}
\email[Corresponding author: ]{alevoj@chalmers.se} 
\affiliation{Department of Applied Physics, Chalmers University of
  Technology, SE-412 96 G\"{o}teborg, Sweden} 

\author{A. Hellman}
\affiliation{Department of Applied Physics, Chalmers University of
  Technology, SE-412 96 G\"{o}teborg, Sweden} 
\affiliation{Competence Centre for Catalysis, Chalmers University of
  Technology, SE-412 96 G\"{o}teborg, Sweden}

\author{C. Ruberto}
\affiliation{Department of Applied Physics, Chalmers University of
  Technology, SE-412 96 G\"{o}teborg, Sweden} 

\author{B.~I. Lundqvist}
\affiliation{Department of Applied Physics, Chalmers University of
  Technology, SE-412 96 G\"{o}teborg, Sweden}
\affiliation{Center for Atomic-scale Materials Design, Department of
  Physics, Technical University of Denmark, DK-2800 Kgs.\ Lyngby,
  Denmark}  

\date{\today}

\begin{abstract}

Adsorption and catalytic properties of the polar (111) surface of
transition-metal carbides (TMC's) are investigated by
density-functional theory. Atomic and molecular adsorption are
rationalized with the concerted-coupling model, in which two types of
TMC surface resonances (SR's) play key roles. The transition-metal
derived SR is found to be a single measurable descriptor for the
adsorption processes, implying that the Br{\o}nsted-Evans-Polanyi
relation and scaling relations apply. This gives a picture with
implications for ligand and vacancy effects and which has a potential
for a broad screening procedure for heterogeneous catalysts.    

\end{abstract}


\maketitle


Finding new simple, efficient, and cheap catalysts is one way to solve
some of today's global environmental challenges \cite{Nocera2007}, as
illustrated by a recent way to harvest solar energy
\cite{Nocera2008}. A goal in catalysis research is to design and tune
the activity and selectivity of catalysts by controlling their
structural properties at the atomic level. This calls for an
identification of key concepts, which can be done from fundamental
theory, in particular density-functional theory (DFT), as shown for
transition-metal (TM) surfaces \cite{Bligaard08}.

For such surfaces, the important role of the TM $d$ electrons for
chemisorption and catalysis was early anticipated \cite{Newns69},
estimated \cite{Lundqvist79}, and articulated in the $d$-band model
\cite{Hammer95, Bligaard08}, which establishes relations between
atomic structure and activity for TM catalysts \cite{Hammer00}. The
$d$-band model explains and predicts a variety of TM properties, for
example, adsorption energies on TM alloys, strengths of bonds at steps
and terraces, and transition-state energies
\cite{Hammer00,Bligaard08}, by correlating them to the energy of the
$d$-band center. The success of the $d$-band model has led to the
introduction of simple descriptors that are able to rationalize
experimental data \cite{Bligaard08}. Examples of such descriptors, 
\textit{e.g.}, for the water gas shift reaction, are the adsorption
energies of oxygen and carbon monoxide on TM surfaces. Today, design of
new TM catalysts by computational screening is a realistic approach
\cite{Sehested2007,Studt}.    

A need for catalysts, a scarcity of precious catalyst materials, and a
general curiosity spur the interest in catalyst materials beyond the
TM's. Statements like ``for several types of reactions, such as
hydrogenation reactions, catalytic activities of carbides and nitrides
were approaching or surpassing those of noble metals'' can be found in
the literature \cite{Chen}. Thus, a theoretical approach extended to
more complex materials than TM's is of interest. In a study of H
adsorption on TM-terminated (111) surfaces of TM carbides (TMC's), a
deviation from the $d$-band model has been observed \cite{Kitchin}. A
recent study of TM oxides, nitrides, and sulfides stresses
similarities in scaling behavior between molecular and atomic
adsorption energies but calls for "a suitably modified $d$-band model" 
\cite{Fernandez}. There exist suggestions of a bulk-derived descriptor
for the reactivity of the TMC's \cite{Toulhoat}, however, our work
goes beyond and finds a surface-derived descriptor based on
electronic-structure calculations. 

Electronic surface states and resonances (SR's), known to appear at
steps and defects, play an important role for the catalytic activity
of a system. The ideal and stable TMC(100) faces are common objects of
study in the literature \cite{Vines} and do not show any presence of
SR's. Recently we have shown that the understanding of atomic
adsorption on TiC and TiN calls for substantial steps beyond the
original $d$-band model \cite{Carlo&Bengt,Alek1,Alek2}. On TiC(111)
and TiN(111), the key actors are identified to be two types of surface
resonances (SR's), derived from Ti (TiSR) and C(N) states (CSR/NSR),
respectively. The adsorption mechanism is explained within a
concerted-coupling model (CCM), in which there are two types of
interactions involved, one between the adsorbate and the TiSR and one
between the adsorbate and the CSR's (NSR's)
\cite{Carlo&Bengt,Alek1,Alek2}. The presence of SR's motivates a focus
on the metastable and polar TMC(111) surfaces, which may, for example,
serve as prototypes for steps and defects.

This Letter demonstrates that an approach that involves DFT
calculations, trend studies, analyses of computed results in
electron-structural terms, formulation of a model, and utilization of
model predictions \cite{Carlo&Bengt,Alek1,Alek2} is useful to reach
the above stated goal. In terms of the CCM, we identify a single
descriptor, the mean energy of the TM-localized SR (TMSR), that
accounts for three important processes on the considered TMC surfaces:
adsorption, dissociation, and catalytic activity. The results should
be general but are here illustrated for one surface reaction, the
ammonia synthesis. As a consequence of the single descriptor,
calculable and experimentally available for some substrates
\cite{Experiment}, scaling relations and Br{\o}nsted-Evans-Polanyi
(BEP) relations are shown to apply, in the same way as for TM surfaces
\cite{Bligaard08}. Finally, the generality and practical usefulness of
the conceptual picture is demonstrated by applications to ligand and
defect effects, which also indicate possibilities for further
refinements. Taken together, these results provide a framework for a
systematic analysis of the catalytic activity of the TMC's.    

Trend studies are conducted for the catalysis-relevant adsorbates H,N,
O, and NH$_x$ ($x=1,2,3$), with respect to the TM component of the
substrate TMC, spanning three different periods and four different
groups in the periodic table (TM = Sc, Ti, V, Zr, Nb, Mo, Ta, and W)
\cite{NaCl}. The first-principles study is performed within the
density-functional theory (DFT), as implemented in DACAPO
\cite{DACAPOCODE,detail1}. The computational procedures are favorably
tested and compared with literature data \cite{ref_calc}. From the DFT
calculations, the adsorption is described in terms of adsorption
energies ($E_{\text{ads}}$) and atom-projected local densities of
states (LDOS).  

\begin{figure}
\centering
\includegraphics[width=0.5\textwidth]{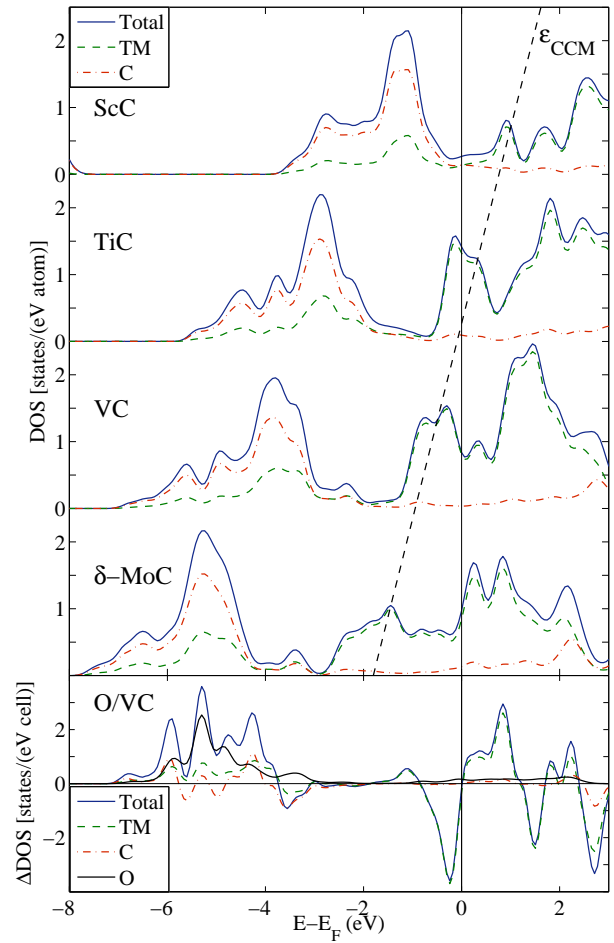}
\caption{\label{fig:DOS}
Upper panel: Calculated atom-projected and total (111) surface DOS's
for a group of TMC's. The dashed diagonal line connects the mean
energies $\varepsilon_{\text{CCM}}$ of the TMSR's and illustrates the
downward shift of the TMSR energy along the TMC series as the group
number of the TM component increases. Lower panel: Difference in
surface DOS induced by O adsorption on VC(111).} 
\end{figure}

Calculated total DOS and LDOS's for the TM-terminated TMC(111)
surfaces (illustrated for some representative cases in Fig.~\ref{fig:DOS}) 
reveal the existence of SR's on all surfaces. There is a TMSR in the  
vicinity of the Fermi energy ($E_{\text{F}}$) and there are several
carbon-localized SR's (CSR's) deeper down in the valence band
\cite{YtAds}. Note that ScC(111) has no filled metallic $d$ SR states.   

The TMSR's are characterized by studying the difference in bulk and
surface DOS's \cite{YtAds}, where they appear as positive peaks owing
to the build up of states at the surface. A parameter
$\varepsilon_{\text{CCM}}$ is defined as the mean energy (center of
gravity) of the TMSR, in close analogy with the $d$-band center
$\varepsilon_{\text{d}}$ in the $d$-band model. The integration is
performed over the energy range of the positive TMSR peak. The value
of $\varepsilon_{\text{CCM}}$ decreases as the group number of the TM
component increases (Fig.~\ref{fig:DOS}), as expected from the filling
of the TM $d$ states \cite{YtAds}.     

Analysis of the DOS before and after adsorption~(illustrated for O/VC
in the lower panel of Fig.~\ref{fig:DOS}) shows that the TMSR is
depleted upon adsorption. This supports the CCM assumption that only
the surface-localized part of the $d$-band spectrum is relevant for
the adsorption and that the TMSR's are key actors in the
adsorbate-surface interaction \cite{YtAds}. Hence,
$\varepsilon_{\text{CCM}}$ is a descriptor for the atomic adsorption
on TMC's. An exception is ScC, with its empty TMSR and consequently an
adsorption mainly arising from the interaction with the CSR's
\cite{YtAds}. This exception confirms the presence of two types of
interactions, as formulated in the CCM
\cite{Carlo&Bengt,Alek1,Alek2,YtAds}.    

\begin{figure}
\centering
\includegraphics[width=0.5\textwidth]{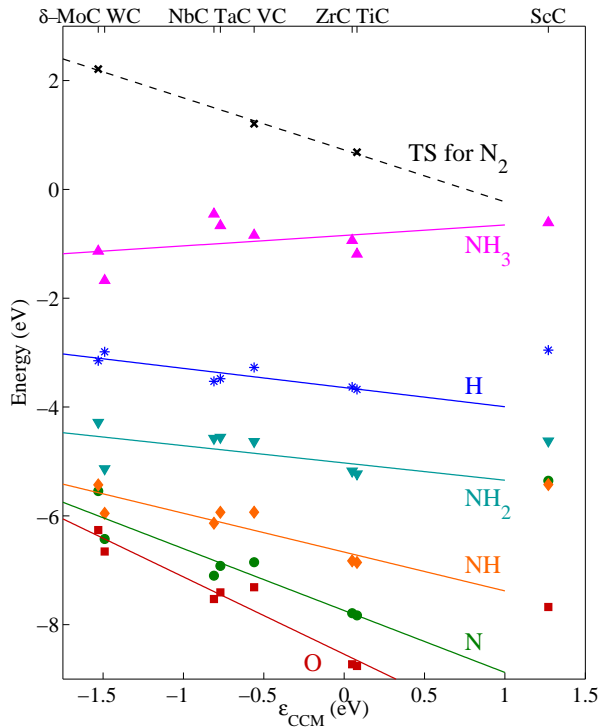}
\caption{\label{fig:Evsed}
Adsorption energies of H, N, O, and NH$_x$ ($x=1,2,3$) on the
considered TMC(111) surfaces \textit{vs.\ }the descriptor
$\varepsilon_{\text{CCM}}$ (the TMSR center of mass). The adsorption
energy is defined as  
$E^{\text{A}}_{\text{ads}} = E_{\text{TMC+A}}-E_{\text{TMC}}-E_{\text{A}}$ 
for each adsorbate A. (ScC is not included in the linear
regressions). Also included (dashed line) are the transition-state
(TS) energies for N$_2$ dissociation, calculated with a fixed
bond-length method.}     
\end{figure}

The role of $\varepsilon_{\text{CCM}}$ as a descriptor is confirmed by
extensive DFT calculations, which yield a linear correlation between
the $E_{\text{ads}}$ and $\varepsilon_{\text{CCM}}$ values for 
each of the studied adsorbates (Fig.~\ref{fig:Evsed}). The
$E_{\text{ads}}$ value decreases (\textit{i.e.}, the adsorption gets
stronger) as $\varepsilon_{\text{CCM}}$ increases, with a gradient
that varies strongly between different adsorbates. The slopes of the
$E_{\text{ads}}$ \textit{vs.}~$\varepsilon_{\text{CCM}}$ correlations
are steeper for the TMC(111) surfaces than for TM's
\cite{Bligaard08}. From a design perspective, this could be a useful
property, allowing larger effects from small changes. Again, ScC does
not follow the trend of the other TMC's, reflecting the different
nature of its interaction with the adsorbate.   

To further test that proper key concepts (TMSR and
$\varepsilon_{\text{CCM}}$) have been identified, ligand effects are
studied. Manipulating the local environment of the surface TM atom
changes $\varepsilon_{\text{CCM}}$, thereby changing adsorption,
activation, and other energies, which opens up for optimization of the
catalytically active site. In our illustrative example, O adsorption
on a TiC(111) surface with one, two, or three surface Ti atoms next to
the fcc adsorption site replaced by V atoms, the calculated adsorption
strength is successively reduced at the same time as the local value
of $\varepsilon_{\text{CCM}}$ is shifted down. This is expected in the
CCM (exchange of Ti with V shifts $d$ levels down in energy) and
monitored by analyzing our calculated LDOS's also in these cases.   

However, there is a deviation from the linear $E_{\text{ads}}$
\textit{vs.\ }$\varepsilon_{\text{CCM}}$ relation \cite{Ligand}. For
example, the $E_{\text{ads}}$ value for the system with three neighboring
V atoms on the TiC(111) surface is similar to that for pure VC(111),
although their $\varepsilon_{\text{CCM}}$ values differ. This ligand
calculation points at further possible refinements of the
descriptor. An orbital projected DOS shows  
that the main interaction takes place between the adsorbate and the
$d_{xz+yz}$ components of the TMSR (the $z$ direction being
perpendicular to the surface). Hence a refined descriptor, taken as
the center of mass of these levels, could be introduced. However, such
a detailed descriptor level might not be practical from an
experimental point of view.

\begin{figure}
\centering
\includegraphics[width=0.5\textwidth]{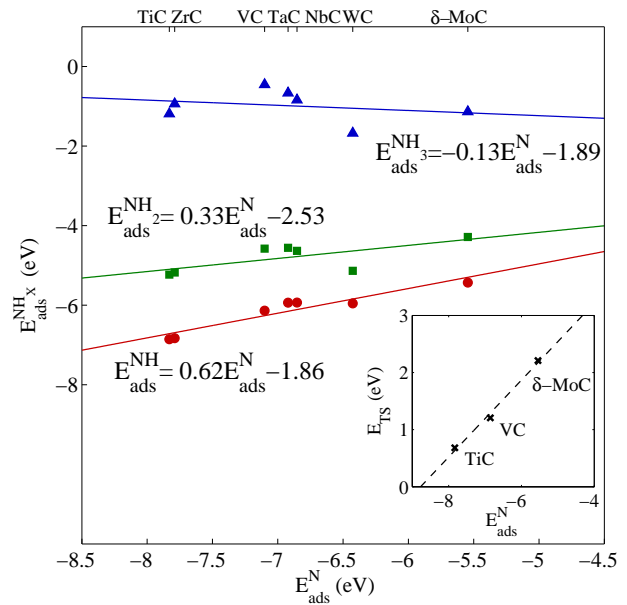}
\caption{\label{fig:scaling}
Adsorption energies on TMC(111) surfaces (ScC is not included) for
NH$_x$ (x$=1,2,3$) \textit{vs.\ }N atomic adsorption energies (scaling
relation). The adsorption energy is defined as
$E^{\text{NH$_x$}}_{\text{ads}}=E_{\text{TMC+NH$_x$}}-E_{\text{TMC}}-E_{\text{NH$_x$}}$. Inset:
the linear variation of the activation energy for N$_2$ dissociation
($E_{\text{TS}}$) \textit{vs.\ }N atomic adsorption energy on TiC, VC,
and $\delta$-MoC (BEP relation).}    
\end{figure}

As $\varepsilon_{\text{CCM}}$ is a descriptor for both atomic and
molecular adsorption (Fig.~\ref{fig:Evsed}), there is a scaling
relation between the molecular and atomic $E_{\text{ads}}$ values
(Fig.~\ref{fig:scaling}). Similar scaling relations have been found
for several molecular species on TM's \cite{Abild-Pedersen} and for
OH$_{\text{x}}$ on TM oxides, NH$_{\text{x}}$ on TM nitrides, and
SH$_{\text{x}}$ on TM sulfides \cite{Fernandez}. These scaling
relations can be described by a linear equation 
$E^{\text{AH}_{x}}_{\text{ads}}=\gamma(x)E^{\text{A}}_{\text{ads}}+\xi$,
where $\gamma(x)$ depends only on the number $x$ of H atoms in the
molecule \cite{Bligaard08}. Our results (Fig.~\ref{fig:scaling}) yield
values ($\gamma_{\text{NH}}=0.62$, $\gamma_{\text{NH}_2}=0.33$,
$\gamma_{\text{NH}_3}= -0.13$) that are close to the predictions
$2/3$, $1/3$, and $0$ for $x = 1$, 2, and 3, respectively. The small
deviations are likely to be caused by the C species in the substrate,
which influences the adsorption mechanism directly via the CSR's and
indirectly via the interaction with the TM atom in the compound
\cite{Kitchin,Carlo&Bengt,Alek1,Alek2}. 

In heterogeneous catalysis, linear correlations (BEP relations)
between activation and adsorption energies play an important role
\cite{Bligaard08}. Activation energies for N$_2$ dissociation on some
TMC(111) surfaces \cite{detail3}, calculated as the transition-state
energy barriers ($E_{\text{TS}}$) by use of a fixed bond-length
method, are presented in the inset of Fig.~\ref{fig:scaling} as a
function of the N adsorption energy. They show that the BEP relation
holds for the TMC's, as expected from the linear correlation between
the activation energies and $\varepsilon_{\text{CCM}}$ (dashed line in
Fig.~\ref{fig:Evsed}). The result is expected to apply generally, as
variations in activation and adsorption energies are governed by the
same basic mechanism, which for TM's is the $d$-band model
\cite{Bligaard08} and for TMC's the CCM
\cite{Carlo&Bengt,Alek1,Alek2,YtAds}.   

The linear dependence of both adsorption and activation energies on
$\varepsilon_{\text{CCM}}$ opens up the possibility to design the TMC
substrates using the CCM to suit different catalytic reactions. This
can be illustrated by the ammonia synthesis (N$_2$+$3$H$_2
\rightleftharpoons 2$NH$_3$), which is a commonly used reaction to
develop new concepts and ideas in catalysis
\cite{HellmanandBoudart}. By using a slightly modified micro-kinetic
model \cite{detail2,Honkala} the calculated catalytic activities of
the TMC's are found to order as a volcano curve with respect to the
single descriptor $\varepsilon_{\text{CCM}}$ \cite{NH3comment}. 

The descriptor $\varepsilon_{\text{CCM}}$ is derived here for TMC's
but is expected to be directly applicable for understanding processes
on other TM compound surfaces (and probably even on other surfaces)
with SR's similar to the ones on the TMC(111) surface. Such SR's can
appear on stable surfaces near defects, such as vacancies, steps, and
adsorbed clusters. For example, experimental studies show that oxygen
atoms replace carbon atoms on TiC(100) and ZrC(100) surfaces when
exposed to O$_2$ \cite{Shirotori,Vines}. Our calculations on an O atom
adsorbed in a C vacancy site on the TiC(100) surface show the presence
of a pronounced TMSR localized around the C vacancy. The calculated
adsorption energy ($-8.71$~eV) is close to the value on the TiC(111)
surface ($-8.76$~eV), to be compared with the value ($-4.96$~eV) on
the vacancy-free TiC(100) surface, where no TMSR is present. This
illustrates the general importance of TMSR's for surface reactions.   

Taken together, our results provide a framework for a systematic
analysis of the catalytic activity of the TMC's. We have shown that by
theoretical means it is possible to find a single descriptor,
corroborated by experimental results on the electronic structure of
the TMC surfaces \cite{Experiment}, for atomic and molecular
adsorption, as well as for a simple reaction on the TMC's. The found
scaling and BEP relations provide simple and efficient connections
between the (local) surface electronic structure and its
reactivity, of importance for studying defect (\textit{e.g.}, vacancy)
and ligand effects. There are thus implications for surfaces,
nanosystems, and catalysis, and the introduced concepts should ramify
to other systems and stimulate the development of similar models for
other classes of materials.        


\textit{Acknowledgments}. The calculations were performed at HPC2N and
NSC via the Swedish National Infrastructure for Computing. Support
from the Swedish Research Council is acknowledged. B.~I. Lundqvist
gratefully acknowledges support from the Lundbeck foundation (Denmark)
via the Center for Atomic-scale Materials Design.   






\begin{thebibliography}{10}
\bibitem{NaCl}
To reduce the complexity and allow for trend studies, all considered
TMC's are described in NaCl structure, which all except WC adopt in
either stable or metastable phase.
\bibitem{detail1}
The spin-polarized approximation of the exchange-correlation
functional according to Perdew-Wang (PW91) \cite{PW91} is
employed. Ultra-soft pseudopotentials \cite{Vanderbilt} and 400~eV
energy cutoff for the one-electron orbitals are used. The $k$-point
sampling of the 3$\times$3$\times$8 TMC(111) slabs is performed in a
(4,4,1) Monkhorst-Pack grid. Repeated slabs are separated by at least
10.8~{\AA} thick vacuum region. During structural relaxations, the two
bottom atomic layers are constrained to the corresponding bulk
geometry. Adsorption is allowed on one (the TM terminated) of the two
slab surfaces in the fcc site (the stable site for adsorption on TiC
\cite{Carlo&Bengt}) and the electrostatic potential is adjusted
accordingly \cite{Bengtsson}. The influence of higher coverage has
been considered in Ref.~\cite{Carlo&Bengt}, illustrating the dominance
of direct adsorbate-substrate forces over lateral interactions.
\bibitem{Ligand}
For the ligand systems with one, two and three V atoms we obtain 
$E_{\text{ads}}=-8.09, -7.78, -7.28$~eV and
$\varepsilon_{\text{CCM}}=0.01, -0.08, -0.11$~eV, respectively.
\bibitem{detail3}
Owing to a strongly adsorbed molecular precursor, relevant activation
energies are calculated with respect to the molecular adsorption energy.
\bibitem{detail2}
It is assumed that N$_2$ dissociation is the rate-determining step and
that all other elementary steps are thus in quasi-equilibrium \cite{Honkala}. 
The turnover frequency (TOF) is given by
TOF$=2kP_{N_2}\theta_*\theta_{N_2}(1-\gamma)$, where $\gamma$ is the
approach to equilibrium. The BEP and scaling relations are used when
calculating the TOF as a function of $\varepsilon_{\text{CCM}}$.
\bibitem{NH3comment}
The ammonia synthesis is used just to prove the benefits of a single
descriptor, not to suggest that TMC's are particularly good catalysts
for it. Rather, reactions requiring a strong molecular precursor, for
instance, oxidation of methane, are a more interesting group of
reactions in a real application
\end{thebibliography}
\end{document}